# Experimental Limits on the Dark Matter Halo of the Galaxy from Gravitational Microlensing

C. Alcock[1,2], R.A. Allsman[3], T.S. Axelrod[1,4], D.P. Bennett[1,2],
K.H. Cook[1,2], K.C. Freeman[4], K. Griest[2,5], J.A. Guern[2,5],
M.J. Lehner[2,5], S.L. Marshall[2,6], H.-S. Park[1], S. Perlmutter[2], B.A. Peterson[4],
M.R. Pratt[2,6], P.J. Quinn[4], A.W. Rodgers[4], C.W. Stubbs[2,6,7], W. Sutherland[2,8]

(The MACHO Collaboration)

1 : Lawrence Livermore National Laboratory, Livermore, CA 94550

2 : Center for Particle Astrophysics, University of California, Berkeley, CA 94720

3 : Supercomputing Facility,
Australian National University, Canberra, A.C.T. 0200, Australia

4 : Mt. Stromlo and Siding Spring Observatories,
Australian National University, Weston, A.C.T. 2611, Australia

5 : Department of Physics, University of California, San Diego, CA 92093

6 : Department of Physics, University of California, Santa Barbara, CA 93106

7 : Departments of Astronomy and Physics, University of Washington, Seattle, WA 98195

8 : Department of Physics, University of Oxford, Oxford OX1 3RH, U.K.

## Abstract

We have monitored 8.6 million stars in the Large Magellanic Cloud for 1.1 years and have found 3 events consistent with gravitational microlensing. We place strong constraints on the Galactic halo content in the form of compact lensing objects in the mass range $10^{-4}\,\mathrm{M}_\odot$ to $10^{-1}\,\mathrm{M}_\odot$. Three events is fewer than expected for a standard spherical halo of objects in this mass range, but appears to exceed the number expected from known Galactic populations. Fitting a naive spherical halo model to our data yields a MACHO fraction $f = 0.20^{+0.33}_{-0.14}$, which implies a total MACHO mass (inside 50 kpc) of $8.0^{+14}_{-6} \times 10^{10}\,\mathrm{M}_\odot$, and a microlensing optical depth $9^{+15}_{-7} \times 10^{-8}$ ($\sim 68\%$ CL).

There is strong evidence from the observed flat rotation curves of spiral galaxies that such galaxies (including our own) have extensive halos of dark matter[1]. We have undertaken an experiment that uses gravitational lensing to search for the dark matter in our own Galactic halo [2]. The experiment is sensitive to lensing by compact objects over a broad range in mass, spanning from $\sim 10^{-7}$ to $\sim 10\,\mathrm{M}_\odot$. This mass regime contains a number of plausible dark matter candidates, including brown dwarfs, black holes, and other stellar remnants. Astrophysical objects of primordial elemental abundances are thought to require a minimum mass of 0.08 $\mathrm{M}_\odot$ to ignite hydrogen fusion, so this experiment is sensitive to non-luminous objects that would have escaped detection in existing sky surveys. Furthermore, we note that a Galactic halo composed entirely of baryonic objects is consistent with the nucleosynthesis bounds on cosmic baryon density [3].

As suggested by Paczynski [4], a population of MAssive Compact Halo Objects (MACHOs) would be detectable by their gravitational 'microlensing' influence on background stars. Microlensing refers to the special case of gravitational lensing in which the splitting of the source star into multiple images is too small to be resolved, but the lensing phenomenon causes a change in the apparent brightness of the source which is time-dependent due to the relative motion of the Earth, lens and source. The apparent amplification $A(t)$ of a background star, in the point-source point-lens approximation, is given by [5]

$$A(t) = A(u(t)) = \frac{u^2+2}{u\sqrt{u^2+4}},$$
$$u(t) = \left[u_{\min}^2 + \left(\frac{2(t-t_{\max})}{\hat{t}}\right)^2\right]^{1/2}, \quad (1)$$

where $\hat{t} = 2r_E/v$ is the time for the lens to move through two Einstein radii ($r_E$) relative to the line of sight, $r_E = \sqrt{4GmD/c^2}$, $m$ is the lens mass, $D = D_{\mathrm{lens}}(D_{\mathrm{source}}-D_{\mathrm{lens}})/D_{\mathrm{source}}$, and $u(t)$ is the distance between the lens and the line of sight in units of $r_E$.

Three teams have reported the detection of gravitational microlensing events, and the world total now exceeds 50 events [2,6-9]. Most of these have been seen in the direction of the Galactic center, but the most suitable target for detecting microlensing by Galactic halo objects is the Large Magellanic Cloud (LMC), located at galactic longitude and latitude $(280^o, -33^o)$ at a distance $D_{source} \approx 50\,\mathrm{kpc}$. For source stars in the LMC the characteristic timescale $\hat{t}$ for microlensing events from a lens of mass $m$ is approximately $130\sqrt{m/\mathrm{M}_\odot}$ days. The fraction of stars being lensed with A $\geq$ 1.34 is defined as the optical depth, $\tau$, which would be [4,10] $\sim 5 \times 10^{-7}$ towards stars in the LMC if the halo is entirely composed of MACHOs.

Since July 1992, we have carried out photometric monitoring of about 8.6 million stars in the LMC using the 1.27-m telescope at Mount Stromlo Observatory, Australia [11].

The data used here consists of our Year 1 data, namely 5169 images of the LMC taken between 21 July 1992 and 3 September 1993, with an exposure time of 300 seconds. The images are distributed between 22 fields near the center of the LMC [12]. The number of images per field ranges from 140 to 350.

Measurements of the relative brightness of the stars are automatically performed on each frame. A good-quality image of each field is first chosen as a 'template', and a list of star positions and magnitudes is generated. All other images are aligned with the template both in position and flux normalization using a set of bright fiducial stars, and a fit is made to the flux of each star, using a point spread function (PSF) determined from the fiducial stars. Each photometric measurement is output with an error estimate and six quality parameters describing the goodness of PSF fit, the flux contamination from nearby stars, and the fractions of flux masked by defective pixels and cosmic-ray events in the CCD array.

The dual passband time series for each star are then searched for events consistent with gravitational microlensing. We use empirically determined cuts on the photometry quality flags, along with the associated data such as seeing and sky brightness for each image, to exclude questionable data points, and we also exclude very red stars which are often irregular variables. The remaining lightcurves are then convolved with a set of filters and any lightcurve showing a peak at a modest significance level is defined as a 'trigger'. For these triggers, a 5-parameter least-squares fit to microlensing is made, defined by $\mathcal{F}_B(t) = A(t)\mathcal{F}_{0B}$ and $\mathcal{F}_R(t) = A(t)\mathcal{F}_{0R}$, where $A(t)$ is given in Eq. 1 above. The fit parameters are the time of peak amplification, $t_{max}$, the characteristic timescale, $\hat{t}$, the peak amplification, $A_{max} = A(u_{min})$, and the baseline flux of the star in red and blue passbands $\mathcal{F}_{0R}$ and $\mathcal{F}_{0B}$. For each trigger, a large set of associated statistics is also computed, and interesting events are automatically selected on the basis of these parameters [13].

There are 4 lightcurves from our year 1 data that pass the above cuts: two of these represent the same star detected independently in a field overlap, thus we have 3 events consistent with microlensing, shown in Fig. 1. The fit parameters are listed in Table 1 below.

| Event | RA (2000) | Dec (2000) | V | V-R | $t_{max}$ (days) | $\hat{t}$ (days) | $A_{max}$ | $\chi^2$ |
|---|---|---|---|---|---|---|---|---|
| 1 | 05 14 44.5 | -68 48 00 | 19.6 | 0.6 | $57.16 \pm 0.02$ | $34.8 \pm 0.2$ | $7.20 \pm 0.09$ | 1.34 |
| 2 | 05 22 57.0 | -70 33 14 | 20.7 | 0.4 | $121.62 \pm 0.3$ | $19.8 \pm 1.3$ | $1.99 \pm 0.06$ | 1.41 |
| 3 | 05 29 37.4 | -70 06 01 | 19.4 | 0.3 | $154.8 \pm 0.9$ | $28.2 \pm 1.7$ | $1.52 \pm 0.03$ | 1.01 |

**Table 1:** Parameters of the events. Columns 4 & 5 show approximate magnitude and color of the lensed stars. Columns 6–8 show the parameters of the best-fit microlensing models: time of peak amplification (Julian days–2449000), the event duration $\hat{t}$, and the peak amplification factor, with the *formal* one sigma errors (derived from the covariance matrix of the fit). Column 9 is the $\chi^2$ per degree of freedom for the microlensing fit.

Quantitative interpretation of our results depends upon the event detection efficiency $\mathcal{E}(\hat{t})$ of the experiment. Inefficiencies arise because of 1) incomplete sampling of the light curves (primarily due to weather interruptions) and 2) because of the blending of two or more stars in an image, only one of which will be lensed. In the case of blending, the overall efficiency is reduced because of distortion of the light curve (and dilution of the amplification); at the same time the number of real target objects is increased, which increases the effective efficiency. These two blending effects partially offset each other.

The efficiency has been determined using a Monte Carlo simulation [13], and the results are shown in Figure 2. The upper curve shows the sampling efficiency (neglecting blending effects), and is an upper bound. The lower curve includes a treatment of the blending efficiency based upon the luminosity function of 'stars' seen in an uncrowded field. The middle curve is based upon a plausible extrapolation of this luminosity function to fainter stars than we can routinely see in the data, and is the 'best estimate' used in this paper.

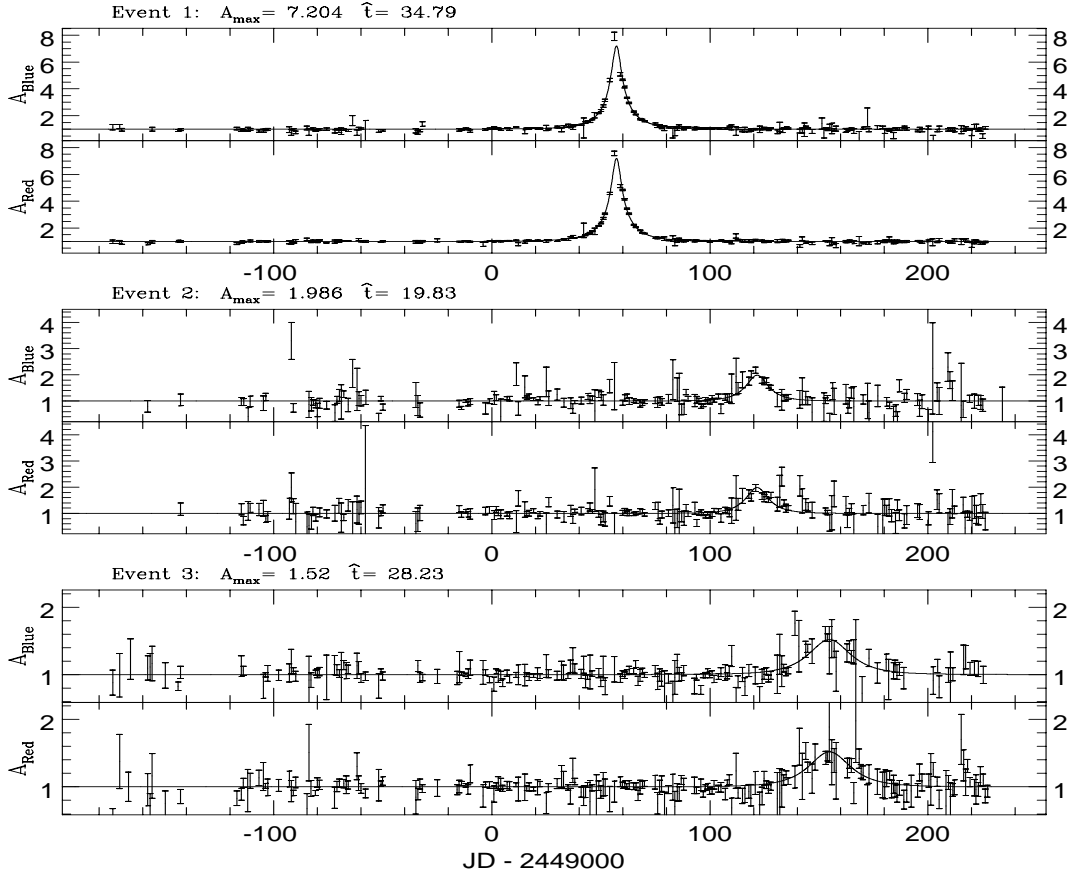

Figure 1: The three observed stellar lightcurves that we interpret as gravitational microlensing events are each shown in relative flux units (red and blue) vs time in days. The solid lines are fits to the theoretical microlensing shape using the parameters given in Table 1.

In order to illustrate the implications of this result we have compared the number and timescales of the observed events with predictions of the commonly used spherical model [1] of the density of dark matter in the Galaxy's halo,

$$\rho_H(r) = \rho_0 \frac{r_0^2 + a^2}{r^2 + a^2}, \qquad (2)$$

where $r$ is galactocentric radius, $r_0 = 8.5\,\mathrm{kpc}$ is the galactocentric radius of the Sun, $a = 5\,\mathrm{kpc}$ is the halo core radius and $\rho_0 = 0.0079\,\mathrm{M_\odot\,pc^{-3}} = 0.30\,\mathrm{GeV\,cm^{-3}}$ is the local dark matter density. This model gives [10] a Galactic halo mass of $4.1 \times 10^{11}\,\mathrm{M_\odot}$ within $50\,\mathrm{kpc}$ of the Galactic center, and predicts a microlensing rate of $\Gamma = 1.6 \times 10^{-6} (m/\mathrm{M_\odot})^{-0.5}$ events/star/yr. After incorporating the detection efficiency, the predicted number of detected events $N_{exp}$ for a dark halo entirely composed of compact objects with a unique mass $m$ is shown in Figure 3. Since we have detected 3 events, models predicting in excess of 7.7 detected events are excluded at 95% confidence. Since we are using only the total number of events, note that exclusion of a delta-function MACHO mass over a given range also applies to *any* arbitrary mass function of MACHOs which is contained within the same range and provides the same total density. Thus, we see from Fig. 3 that lensing objects in the mass range $8 \times 10^{-5}\,M_\odot < m < 0.3\,M_\odot$ comprise $< 100\%$ of the model halo at 95% CL.

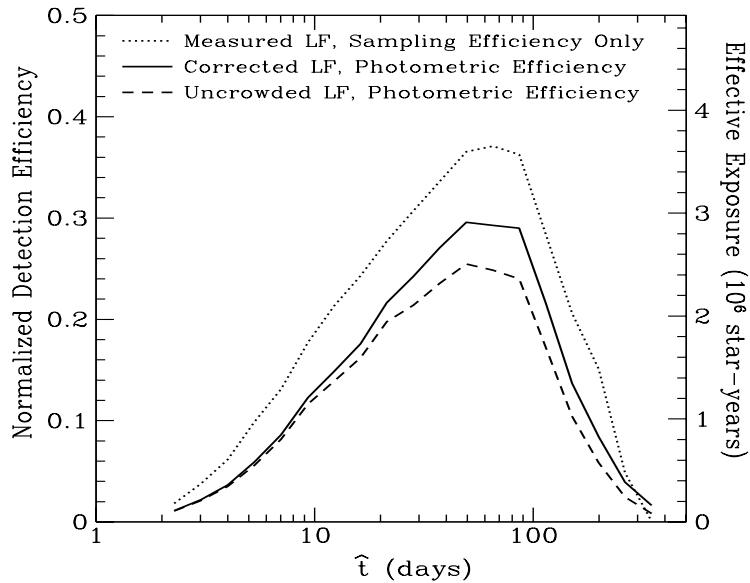

Figure 2: MACHO Year 1 microlensing event detection efficiency. Here $\mathcal{E}$ is shown as a function of $\hat{t}$, the event duration. The left axis is labeled with fractional event detection efficiency per monitored object, and the right axis gives the efficiency–corrected exposure. The upper curve takes into account only the temporal sampling effects while the lower two correct for event degradation from blending by using two different stellar luminosity functions (LFs) to estimate the increased number of targets. The best estimate is the middle curve.

Current estimates [14] of the Galactic mass within 50 kpc range from 4 to $6 \times 10^{11}\,M_\odot$, while the model of equation 2 contains $4.1 \times 10^{11}\,M_\odot$. Using the 95% CL limits on $N_{exp}$ we can set a 95% CL upper limit on the mass in this halo of $M_{lim} = 4.1 \times 10^{11}\,M_\odot (7.7/N_{exp})$. As shown in Figure 3, objects in the mass range $3 \times 10^{-4}\,M_\odot < m < 0.06\,M_\odot$ contribute no more than 50% of the model halo's mass within 50 kpc.

We have explored a range of different halo density profiles [13], and we find that while the constraint on the halo mass *fraction* in MACHOs is quite model-dependent, our constraints on the *total* mass of MACHOs interior to 50 kpc are relatively independent of the assumed model of the galactic halo.

In setting the limits above we need not assume that our three events are due to microlensing by halo objects; however, if we add this assumption we can use the observed durations of the events in a maximum likelihood analysis to find the most likely MACHO fraction $f$, and mass $m$ (again with a delta function mass distribution). In this model the rest of the halo is presumed to consist of objects with masses outside our range of sensitivity. Figure 4 shows the likelihood contours with most likely values of $f = 0.20^{+0.33}_{-0.14}$ and $m = 0.06^{+0.11}_{-0.04}$ M$_\odot$ (errors given are the total extent of the 68% contours).

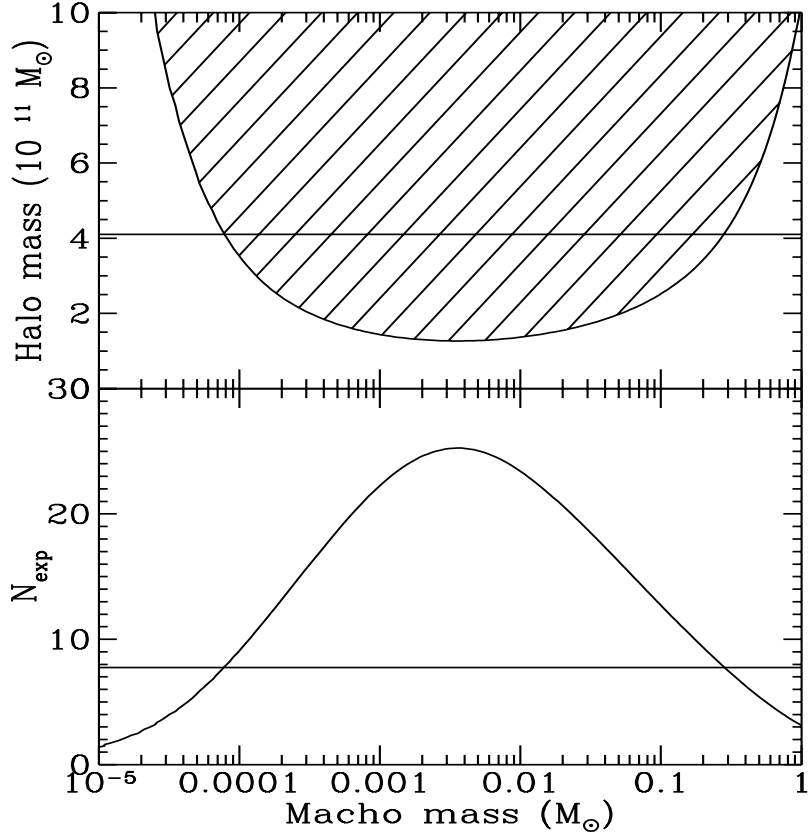

Figure 3: The lower panel shows the number of expected events predicted from the standard model halo with a delta function mass distribution. Given three observed events, points above the line drawn at $N_{exp} = 7.7$ are excluded at the 95% CL. The upper panel shows the 95% CL limit on the halo mass in MACHOs within 50 kpc of the galactic center for the model. Points above the curve are excluded at 95% CL while the line at $4.1 \times 10^{11}$ M$_\odot$ shows the total mass in this model within 50 kpc.

The model of equation 2 predicts an optical depth of $\tau_{model} = 4.7 \times 10^{-7}$. Using $f$ we can now state the model best fit optical depth of $9.2^{+15}_{-6.7} \times 10^{-8}$ and observed MACHO mass within 50 kpc of $8.0^{+14}_{-6} \times 10^{10}$ M$_\odot$. We can also estimate the optical depth directly using our observed timescales, our exposure $E = 9.72 \times 10^6$ star years, and efficiencies by computing $\tau_{est} = \frac{\pi}{4E} \Sigma(\hat{t}_i/\mathcal{E}_i) = 8.0 \times 10^{-8}$, in good agreement with the maximum likelihood estimate.

For three detected events the 90% CL lower bound on the underlying rate is 1.1 events. We have estimated [13] the event rates due to the galactic disk, the spheroid and the LMC disk, and find that they should contribute on average $\sim \frac{1}{2}$ an event to our sample. Our continuing observations should help to ascertain the significance of this difference. We caution that recent microlensing results toward the Galactic bulge suggest that the standard galactic models used here may be incorrect [7-9].

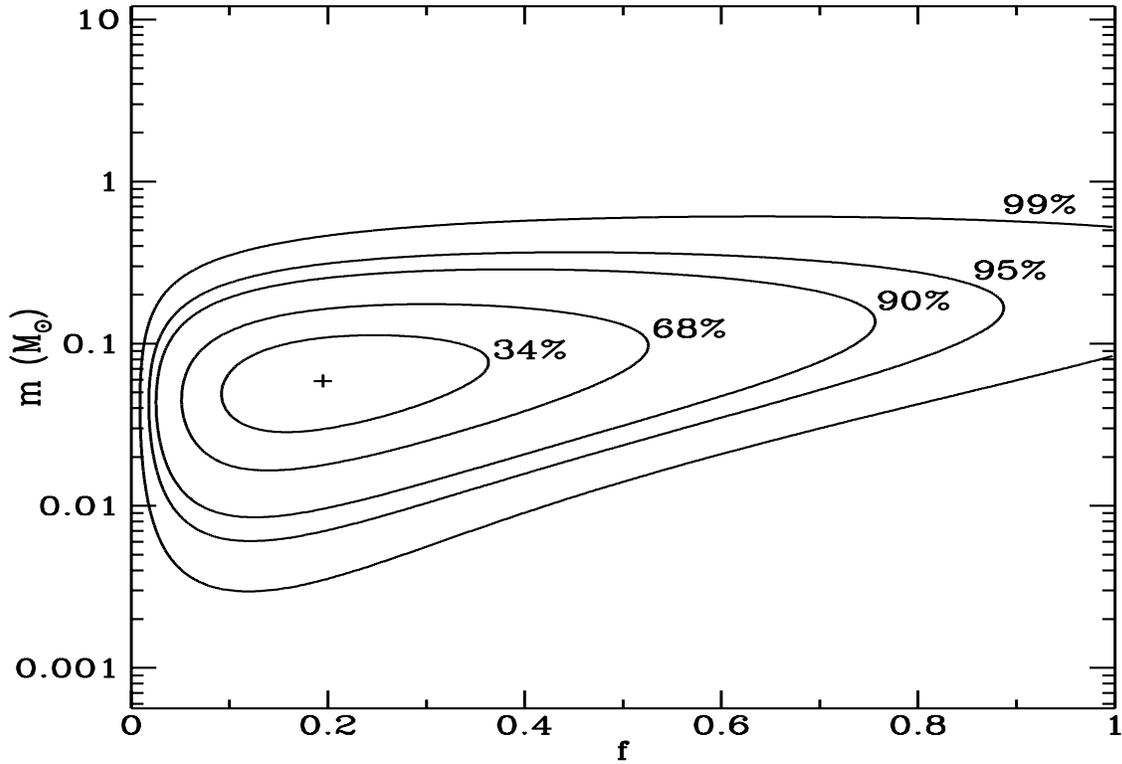

Figure 4: Likelihood contours for the model halo with a delta function mass distribution. The numbers labeling the contour levels indicate the total probability of the enclosed parameter region according to a Bayesian analysis assuming a prior distribution uniform in $m$ and $f$. The position of the most likely MACHO mass $m$ and MACHO halo fraction $f$ is marked with a $+$.

We are grateful for the support given our project by the technical staff at the Mt. Stromlo Observatory. Work performed at LLNL is supported by the DOE under contract W-7405-ENG. Work performed by the Center for Particle Astrophysics personnel is supported by the NSF through AST 9120005. The work at MSSSO is supported by the Australian Department of Industry, Science and Technology. K.G. acknowledges a DOE OJI grant. C.S. acknowledges the generous support of the Packard Foundation, and both K.G. and C.S are grateful to the Sloan Foundation for their support.